\documentclass[epj]{svjour}
\usepackage{graphics}
\newcommand{\al}{$\alpha$}
\newcommand{\ald}{$\alpha$-decay}

\newcommand{\thcalc}{$T_{1/2,\alpha}^{\rm{calc}}$}

\newcommand{\thpre}{$T_{1/2,\alpha}^{\rm{pre}}$}
\begin{document}
\title{
Super-allowed $\alpha$ decay above doubly-magic $^{100}$Sn and
properties of $^{104}$Te = $^{100}$Sn $\otimes$ $\alpha$
}
\author{
Peter Mohr\inst{1} \thanks{\emph{email:} WidmaierMohr@compuserve.de}
}                     
%
%
\institute{
Diakoniekrankenhaus, D-74523 Schw\"abisch Hall, Germany
}
\date{Received: date / Revised version: date}
%
\abstract{ 
\ald\ half-lives for $^{104,105,106}$Te and $^{108,109,110}$Xe close
above the doubly-magic $^{100}$Sn are calculated from systematic
double-folding potentials. The derived \al\ preformation factors are
compared to results for $^{212,213,214}$Po and $^{216,217,218}$Rn
above the doubly-magic $^{208}$Pb. \ald\ energies of $Q_\alpha = 5.42
\pm 0.07$\,MeV and $4.65 \pm 0.15$\,MeV are predicted for $^{104}$Te
and $^{108}$Xe; the corresponding half-lives are $T_{1/2} \approx
5$\,ns for $^{104}$Te and of the order of $60\,\mu$s for $^{108}$Xe.
Additionally, the properties of
rotational bands in $^{104}$Te are analyzed, and the first excited
$2^+$ state in $^{104}$Te is predicted at $E_x = 650 \pm 40$\,keV; it
decays preferentially by $\gamma$ emission with
a reduced transition strength of 10 Weisskopf units to the
ground state of $^{104}$Te and with a minor branch by \al\ emission to
the ground state of $^{100}$Sn.
} 
\titlerunning{
Super-allowed $\alpha$ decay above doubly-magic $^{100}$Sn and
properties of $^{104}$Te = $^{100}$Sn $\otimes$ $\alpha$
}
\maketitle
\section{Introduction}
\label{intro}
Studies of \ald\ properties of nuclei with $Z \approx N$ in the mass
region above $A \approx 100$ have been stimulated by recent
experimental progress: Seweryniak {\it et al.}\ \cite{Sew06} have
detected the \ald\ of $^{105}$Te at the Argonne Fragment Analyzer, and
Liddick {\it et al.}\ \cite{Lid06} have analzed the \ald\ chain
$^{109}$Xe(\al )$^{105}$Te(\al )$^{101}$Sn at the Recoil Mass
Spectrometer of the Holifield Radioactive Ion Beam Facility. In both
papers the measured \ald\ half-lives are interpreted as indication for
super-allowed \ald\ in the vicinity of the doubly-magic nucleus
$^{100}$Sn with $Z = N = 50$. Whereas the larger experimental
uncertainties in \cite{Sew06} allowed only to conclude ``a modest
enhancement of \ald\ rates toward the $N=Z$ line'', the latest data of
\cite{Lid06} clearly confirm the super-allowed \ald\ of $^{105}$Te by
comparison with the analogous \ald\ of $^{213}$Po.
A first theoretical report by Xu and Ren
\cite{Xu06} is based on improved folding potentials, and they find an
increased \al\ preformation factor for $N = Z$ nuclei.

The present study reanalyzes the new experimental data
\cite{Sew06,Lid06} using a similar model as \cite{Xu06} in combination
with double-folding potentials which are close to the results of
elastic scattering data on $N \approx Z$ data in the $A \approx 100$
mass region ($^{92}$Mo \cite{Ful01}, $^{106}$Cd \cite{Kis06}, $^{112}$Sn
\cite{Gal05}). The results further confirm the super-allowed \ald\
around $^{100}$Sn. The systematic properties of the double-folding
potentials allow a prediction of the \ald\ energy of $^{104}$Te and
$^{108}$Xe with relatively small uncertainties. However, the
prediction of the \ald\ half-lives has still considerable
uncertainties because of the exponential dependence on the \ald\
energy. In addition, \al\ cluster properties of the nucleus $^{104}$Te
= $^{100}$Sn $\otimes$ \al\ can be predicted in a similar way as in
\cite{Ohk95,Mic98} for $^{94}$Mo = $^{90}$Zr $\otimes$ \al
. In particular, the excitation energy of the first excited $2^+$ state
in $^{104}$Te and its decay properties by $\gamma$ and \al\ emission
are calculated. These decay properties have noticeable influence on the
experimental determination of the \ald\ of $^{104}$Te.

\section{\ald\ half-lives}
\label{sec:half}
In a semi-classical approximation the $\alpha$-decay width
$\Gamma_\alpha$ is given by the following formulae \cite{Gur87}:
\begin{equation}
\Gamma_\alpha = P F \frac{\hbar^2}{4\mu} 
\exp{\left[ -2 \int_{r_2}^{r_3} k(r) dr \right]}
\label{eq:gamma}
\end{equation}
with the preformation factor $P$, the normalization factor $F$ 
\begin{equation}
F \int_{r_1}^{r_2} \frac{dr}{2\,k(r)} = 1
\label{eq:f}
\end{equation}
and the wave number $k(r)$
\begin{equation}
k(r) = \sqrt{ \frac{2\mu}{\hbar^2}\left|E - V(r)\right|} \quad \quad .
\label{eq:k}
\end{equation}
$\mu$ is the reduced mass and $E$ is the decay energy of the
$\alpha$-decay which was taken from the mass table of
Ref.~\cite{Wap03} and the recent experimental results of
\cite{Sew06,Lid06}. The $r_i$ are the classical turning points. For
$0^+ \rightarrow 0^+$ $s$-wave decay the inner turning point is at
$r_1 = 0$. $r_2$ varies around 7\,fm, and $r_3$ varies strongly
depending on the energy. The decay width $\Gamma_\alpha$ is related to
the half-life by the well-known relation $\Gamma_\alpha = \hbar \ln{2}
/ T_{{1/2},\alpha}$. Following Eq.~(\ref{eq:gamma}), the preformation
factor may also be obtained as
\begin{equation}
P = \frac{ T_{1/2,\alpha}^{\rm{calc}} }{ T_{1/2,\alpha}^{\rm{exp}} }
\label{eq:pre}
\end{equation}
where $\Gamma_\alpha$ or \thcalc\ are calculated from
Eq.~(\ref{eq:gamma}) with $P = 1$. For completeness, I define the here
predicted half-life for unknown nuclei as \thpre $ = $\thcalc
$/P$. Further details of the model can be found in
\cite{Mohr00,Mohr06}. 

The potential $V(r)$ in Eq.~(\ref{eq:k}) is given by 
\begin{equation}
V(r) = V_N(r) + V_C(r) = \lambda \, V_F(r) + V_C(r)
\label{eq:vtot}
\end{equation}
where the nuclear potential $V_N$ is the double-folding potential
$V_F$ multiplied by a strength parameter $\lambda \approx 1.1 - 1.3$
\cite{Atz96}. The nuclear densities have been taken from \cite{Vri87}
in the same parametrization as in \cite{Mohr06} for all nuclei under study.
$V_C$ is the Coulomb potential in the usual form of a
homogeneously charged sphere with the Coulomb radius $R_C$ chosen
the same as the $rms$ radius of the folding potential $V_F$. For
decays with angular momenta $L \ne 0$ an additional centrifugal
potential $V_L = L(L+1) \hbar^2/(2 \mu r^2)$ is used.

The potential strength parameter $\lambda$ of the folding potential
was adjusted to the 
energy of the $\alpha$ particle in the $\alpha$ emitter
$(A+4) = A \otimes \alpha$. The number of nodes of the bound state
wave function was taken from the Wildermuth condition
\begin{equation}
Q = 2N + L = \sum_{i=1}^4 (2n_i + l_i) = \sum_{i=1}^4 q_i
\label{eq:wild}
\end{equation}
where $Q$ is the number of oscillator quanta,
$N$ is the number of nodes and $L$ the relative angular
momentum of the $\alpha$-core wave function, and
$q_i = 2n_i + l_i$ are the corresponding quantum numbers
of the nucleons in the $\alpha$ cluster. I have taken
$q = 4$ for $50 < Z,N \le 82$, 
$q = 5$ for $82 < Z,N \le 126$ and
$q = 6$ for $N > 126$
where $Z$ and $N$ are the proton and neutron number of the daughter
nucleus. This leads to $Q = 16$ for the nuclei above $^{100}$Sn and $Q
= 22$ for the nuclei above $^{208}$Pb.

\begin{figure}
\resizebox{0.5\textwidth}{!}{
  \includegraphics[65,55][470,345]{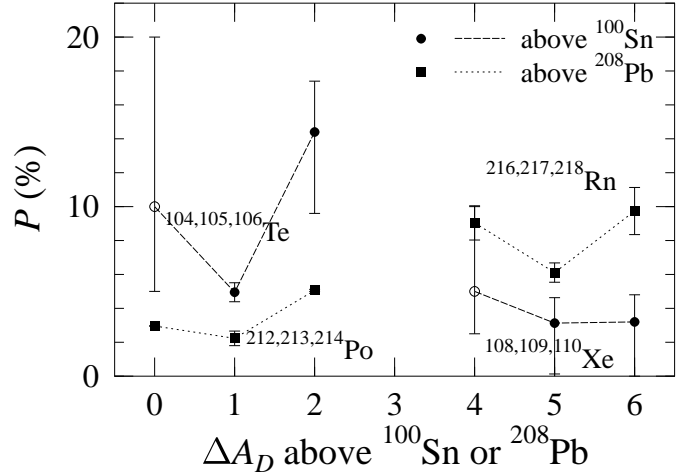}
}
\caption{
Comparison of preformation factors $P$ for the \ald s of
$^{104,105,106}$Te and $^{108,109,110}$Xe above doubly-magic
$^{100}$Sn (circles) and $^{212,213,214}$Po and $^{216,217,218}$Rn
above doubly-magic $^{208}$Pb (squares), derived from
Eq.~(\ref{eq:pre}). The open circles for $^{104}$Te and $^{108}$Xe
indicate assumed values: $P = 10$\,\% for $^{104}$Te and $P = 5$\,\%
for $^{108}$Xe. The lines are to guide the eye only.
}
\label{fig:pre}
\end{figure}

The results for the nuclei $^{108,109,110}$Xe and $^{104,105,106}$Te
above the doubly-magic $^{100}$Sn and for $^{216,217,218}$Rn and
$^{212,213,214}$Po above the doubly-magic $^{208}$Pb are listed in
Table \ref{tab:halflife}. The derived preformation factors $P$ are
shown in Fig.~\ref{fig:pre} as a function of $\Delta A_D$ where
$\Delta A_D$ gives the distance from a double shell closure. E.g., the
preformation factor for the \ald\ $^{106}$Te $\rightarrow$ $^{102}$Sn
can be found at $\Delta A_D = 2$ because the daughter nucleus
$^{102}$Sn has two nucleons above the doubly-magic $^{100}$Sn. The
same value of $\Delta A_D = 2$ is found for the \ald\ $^{214}$Po
$\rightarrow$ $^{210}$Pb. Thus, a comparison between the results above
$A = 100$ and above $A = 208$ can be done easily.

\begin{table*}
\setlength{\tabcolsep}{1.5ex}
\caption{
\ald\ half-lives for nuclei above $^{100}$Sn and $^{208}$Pb.
}
\label{tab:halflife}       
\begin{tabular}{r@{$\rightarrow$}lr@{$\rightarrow$}lllclll}
\hline\noalign{\smallskip}
\multicolumn{2}{c}{decay} & $J_i$ & $J_f$ & \multicolumn{1}{c}{$E$} &
\multicolumn{1}{c}{$\lambda$} & \multicolumn{1}{c}{$J_R$} &
\multicolumn{1}{c}{$T_{1/2}^{\rm{exp}}$ or $T_{1/2}^{\rm{pre}}$} & 
\multicolumn{1}{c}{$T_{1/2}^{\rm{calc}}$} & \multicolumn{1}{c}{$P$} \\ 
\multicolumn{2}{c}{} & \multicolumn{2}{c}{} &
\multicolumn{1}{c}{(MeV)} & & \multicolumn{1}{c}{(MeV\,fm$^3$)}
& \multicolumn{1}{c}{(s)} & \multicolumn{1}{c}{(s)} &
\multicolumn{1}{c}{(\%)} \\ 
\noalign{\smallskip}\hline\noalign{\smallskip}
$^{218}$Rn & $^{214}$Po & $0^+$ & $0^+$ & 7.263 & 1.2431 & 328.2 &
$(3.5 \pm 0.5) \times 10^{-2}$ & $3.41 \times 10^{-3}$ & 
$9.74 \pm 1.39$ \\ 
$^{217}$Rn & $^{213}$Po & $9/2^+$ & $9/2^+$ & 7.887 & 1.2390 & 327.2 &
$(5.4 \pm 0.5) \times 10^{-4}$ & $3.30 \times 10^{-5}$ & 
$6.11 \pm 0.57$ \\ 
$^{216}$Rn & $^{212}$Po & $0^+$ & $0^+$ & 8.200 & 1.2386 & 327.2 &
$(4.5 \pm 0.5) \times 10^{-5}$ & $4.07 \times 10^{-6}$ & 
$9.04 \pm 1.01$ \\ 
$^{214}$Po & $^{210}$Pb & $0^+$ & $0^+$ & 7.834 & 1.2384 & 327.3 &
$(1.64 \pm 0.02) \times 10^{-4}$ & $8.32 \times 10^{-6}$ & 
$5.06 \pm 0.06$ \\ 
$^{213}$Po & $^{209}$Pb & $9/2^+$ & $9/2^+$ & 8.536 & 1.2333 & 326.1 &
$(4.2 \pm 0.8) \times 10^{-6}$ & $9.38 \times 10^{-8}$ & 
$2.23 \pm 0.43$ \\ 
$^{212}$Po & $^{208}$Pb & $0^+$ & $0^+$ & 8.954 & 1.2316 & 325.7 &
$(2.99 \pm 0.02) \times 10^{-7}$ & $8.70 \times 10^{-9}$ & 
$2.96 \pm 0.02$ \\ 
$^{110}$Xe & $^{106}$Te & $0^+$ & $0^+$ & 3.885 & 1.0981 & 302.4 &
$\approx 4 \times 10^{-1}$ $^a$ & $1.29 \times 10^{-2}$ & 
$\approx 3.2$ \\ 
$^{109}$Xe & $^{105}$Te & $7/2^+$ & $7/2^+$ & 4.067 & 1.1006 & 303.2 &
$(1.3 \pm 0.2) \times 10^{-2}$ & $1.42 \times 10^{-3}$ $^b$ & 
$\approx 3$ $^b$ \\ 
$^{108}$Xe & $^{104}$Te & $0^+$ & $0^+$ & 4.65 $^c$ & 1.099 & 303.4 &
$\approx 60$\,$\mu$s $^{c,d}$ & $\approx 3 \times 10^{-6}$ $^d$ & 
$\approx 5$ $^e$ \\ 
$^{106}$Te & $^{102}$Sn & $0^+$ & $0^+$ & 4.290 & 1.1026 & 304.5 &
$(6.0^{+3.0}_{-1.0}) \times 10^{-5}$ & $8.66 \times 10^{-6}$ & 
$14.4^{+3.0}_{-4.8}$ \\ 
$^{105}$Te & $^{101}$Sn & $5/2^+$ & $5/2^+$ & 4.889 & 1.1006 & 304.1 &
$(6.2 \pm 0.7) \times 10^{-7}$ & $3.07 \times 10^{-8}$ & 
$4.95 \pm 0.56$ \\ 
$^{104}$Te & $^{100}$Sn& $0^+$ & $0^+$ & 5.42 $^c$ & 1.100 & 304.0 &
$\approx 5$\,ns $^c$ & $\approx 5 \times 10^{-10}$ & 
$\approx 10$ $^e$ \\
\noalign{\smallskip}\hline
\multicolumn{8}{l}{$^a$ \ald\ branch only} \\
\multicolumn{8}{l}{$^b$ branching to $7/2^+$: see Sect.~\ref{sec:half}} \\
\multicolumn{8}{l}{$^c$ predicted values; see Sect.~\ref{sec:predict}} \\
\multicolumn{8}{l}{$^d$ huge uncertainty from unknown energy $E$; see
  Sect.~\ref{sec:predict}} \\ 
\multicolumn{8}{l}{$^e$ assumed values; see Fig.~\ref{fig:pre}} \\
\end{tabular}
\end{table*}

The systematic behavior of the potential parameters is one main
advantage of the folding potentials. The potential strength parameter
$\lambda$ and the normalized volume integral per interacting nucleon
pair
\begin{equation}
J_R = \frac{\lambda}{A_P A_T} \, \int V_F(r) \, d^3r
\label{eq:jr}
\end{equation}
show values around $\lambda \approx 1.10$ and $J_R \approx
303$\,MeV\,fm$^3$ for the systems $^{100,101,102}$Sn $\otimes$ \al\
and $^{104,105,106}$Te $\otimes$ \al\ above $A = 100$; the
variations of $\lambda$ and $J_R$ are less than 1\,\% and allow thus
extrapolations with limited uncertainties. The same range of
variations of less than 1\,\% is found for the considered systems
above $A = 208$ where $\lambda \approx 1.24$ and $J_R \approx
327$\,MeV\,fm$^3$.

The analysis of the $0^+ \rightarrow 0^+$ decays of the even-even
systems is straightforward. The ground state transitions
dominate because these transitions have the maximum energy, and the
decay is not hindered by an additional centrifugal barrier because $L
= 0$. In both decays $^{217}$Rn $\rightarrow$ 
$^{213}$Po and $^{213}$Po $\rightarrow$ $^{209}$Pb the ground state
transitions $9/2^+ \rightarrow 9/2^+$ with $L = 0$ also
dominate. However, the analysis of the \ald s $^{109}$Xe $\rightarrow$
$^{105}$Te and $^{105}$Te $\rightarrow$ $^{101}$Sn requires further
study. 

Two \al\ groups have been detected in the decay of $^{109}$Xe
$\rightarrow$ $^{105}$Te which have been interpreted as the $L = 0$
and $L = 2$ decays from the $7/2^+$ ground state of $^{109}$Xe to the
$5/2^+$ ground state and $7/2^+$ first excited state in $^{105}$Te
\cite{Lid06}. From Eq.~(\ref{eq:gamma}) one calculates \thcalc\ $=
5.71 \times 10^{-4}$\,s for the $L = 2$ ground state decay and
\thcalc\ $= 1.42 \times 10^{-3}$\,s for the $L = 0$ decay to the first
excited state, in both cases using $P = 1$. The theoretical branching
is 71\,\% for the ground state branch and 29\,\% for the branch to the
first excited state. This is in excellent agreement with the
experimental values of $(70 \pm 6)$\,\% for the ground state branch
and $(30 \pm 6)$\,\% for the branch to the first excited state
\cite{Lid06}. In Fig.~\ref{fig:pre} I show the preformation factor $P$
in $^{109}$Xe for the $L = 0$ decay only because all the other decays
in Fig.~\ref{fig:pre} have the same $L = 0$.

For the \ald\ $^{105}$Te $\rightarrow$ $^{101}$Sn only one \al\ group
has been detected in \cite{Lid06}, and an upper limit of 5\,\% is
given for other decay branches. The \ald\ strength increases with
increasing energy and decreasing angular momentum. If only one decay
branch is observed, one may conclude that this branch corresponds to
a $L = 0$ ground state transition. Consequently,
$J^\pi(^{101}{\rm{Sn}}) = J^\pi(^{105}{\rm{Te}}) = 5/2^+$
\cite{Lid06}. This is in agreement with a recent theoretical
prediction \cite{Kav06}. The derived values for the potential strength
parameter $\lambda$ and the volume integral $J_R$ fit into the
systematics and thus strengthen the above tentative spin assignment.

The results in Fig.~\ref{fig:pre} and Table \ref{tab:halflife} confirm
the super-allowed nature of \ald\ near the doubly-magic
$^{100}$Sn. For $^{216,217,218}$Rn one finds preformation values $P$
between about 5\,\% and 10\,\%. Surprisingly, $P$ slightly decreases
for $^{212,213,214}$Po to values between about 2\,\% and 5\,\% when
approaching the doubly-magic daughter nucleus $^{208}$Pb. For
$^{109,110}$Xe relatively small values of $P \approx 3\,\%$ are
found. When approaching the doubly-magic daughter $^{100}$Sn, the
preformation values $P$ show the expected behavior and increase to
about 5\,\% to 15\,\% for $^{105,106}$Te. A comparison between the
preformation factors $P$ for the Po isotopes and the Te isotopes shows
that 
\begin{equation}
P({\rm{Te}}) \approx 3 \times P({\rm{Po}})
\label{eq:precomp}
\end{equation}
in agreement with the conclusions of \cite{Sew06,Lid06}.

\section{Predicted half-lives of $^{104}$Te and $^{108}$Xe}
\label{sec:predict}
The systematic behavior of the potential parameters $\lambda$ and
$J_R$ in combination with the shown preformation factors $P$ (see
Fig.~\ref{fig:pre}) enables the extrapolation to the decays $^{108}$Xe
$\rightarrow$ $^{104}$Te $\rightarrow$ $^{100}$Sn with limited
uncertainties. For the prediction of the \ald\ energies I use a local
potential which is adjusted to the neighboring nuclei. 

The potentials for $^{105}$Te = $^{101}$Sn $\otimes$ \al\ and
$^{106}$Te = $^{102}$Sn $\otimes$ \al\ are practically identical. From
the average $J_R = 304.29$\,MeV\,fm$^3$ one obtains the \ald\ energy
of $^{104}$Te $E = 5.354$\,MeV, whereas a linear extrapolation yields
a slighly weaker potential $J_R = 303.76$\,MeV\,fm$^3$ and slightly
higher energy $E = 5.481$\,MeV. Combining these results, a reasonable
prediction of the \ald\ energy of $^{104}$Te is $E = 5.42 \pm
0.07$\,MeV. 

From the lower decay energy $E = 5.354$\,MeV one obtains \thcalc\ $=
7.87 \times 10^{-10}$\,s from Eq.~(\ref{eq:gamma}) with $P = 1$; the
higher decay energy yields \thcalc\ $= 3.13 \times 10^{-10}$\,s. The
uncertainty of the \ald\ energy of about 70\,keV translates to an
uncertainty in the calculated half-life of about a factor of 1.5. For
a prediction of the \ald\ half-life one has to find a reasonable
assumption for the preformation factor $P$. Following the pattern of
$P$ in Fig.~\ref{fig:pre}, I use $P = 10$\,\% with an estimated
uncertainty of a factor of two. Combining the above findings, the
predicted half-life of $^{104}$Te is \thpre\ $= 5$\,ns with an
uncertainty of about a factor three. The uncertainty of the predicted
half-life is composed of similar contributions for the unknown \ald\
energy and the assumed preformation factor $P$.

The potentials for $^{109}$Xe = $^{105}$Te $\otimes$ \al\ and
$^{110}$Xe = $^{106}$Te $\otimes$ \al\ change by about 1\,MeV\,fm$^3$;
this is still very similar, but not as close as in the above
$^{105}$Te = $^{101}$Sn $\otimes$ \al\ and $^{106}$Te = $^{102}$Sn
$\otimes$ \al\ cases. Repeating the above procedure, one finds the
\ald\ energy $E = 4.792$\,MeV from the average $J_R =
302.82$\,MeV\,fm$^3$ and $E = 4.506$\,MeV from the extrapolated $J_R =
303.96$\,MeV\,fm$^3$. The calculated half-lives using $P = 1$ are
\thcalc\ $= 7.40 \times 10^{-7}$\,s for the higher energy $E =
4.792$\,MeV and \thcalc\ $= 1.18 \times 10^{-5}$\,s for the lower
energy $E = 4.506$\,MeV. Combining these results, the \ald\ energy is
$E = 4.65 \pm 0.15$\,MeV. Together with a preformation factor of
about $P = 5$\,\% the \ald\ half-life is predicted to be of the order
of 100\,$\mu$s. However, the uncertainty of the decay energy of
150\,keV leads to an uncertainty in the half-life of a factor of 4;
thus it is impossible to predict the \ald\ half-life of $^{108}$Xe
better than this uncertainty.

\begin{figure}
\resizebox{0.5\textwidth}{!}{
  \includegraphics[40,60][470,410]{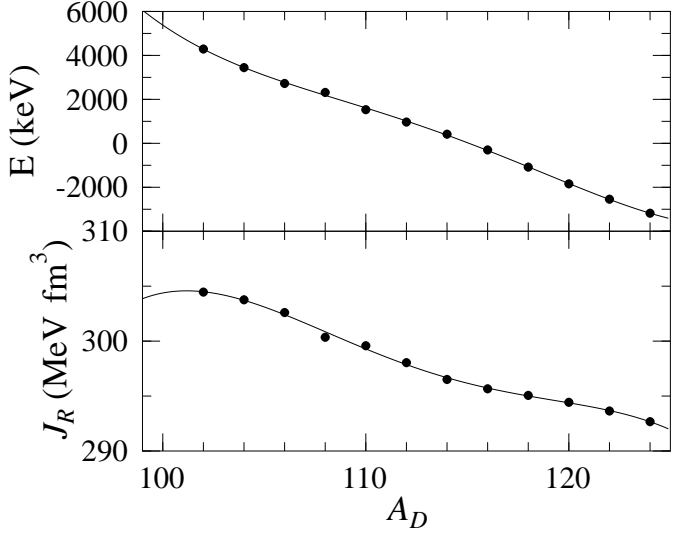}
}
\caption{
Volume integral $J_R$ and energy $E$ in dependence of the mass number
$A_D$ from $^{102}$Sn to $^{124}$Sn (See text).
}
\label{fig:a100fit}
\end{figure}

It is interesting to compare the predictions for the \ald\ properties
of $^{104}$Te with the results of \cite{Xu06}. In \cite{Xu06} the
\ald\ energy is linearly extrapolated from the neighboring even-even
Te isotopes $^{106,108,110}$Te leading to $E = 5.053$\,MeV. I have
repeated this procedure for the Te isotopes $^{106}$Te to
$^{126}$Te. The \ald\ energies and derived volume integrals $J_R$ are
shown in Fig.~\ref{fig:a100fit}.
For an extrapolation to the \ald\ of $^{104}$Te I have fitted the data
in Fig.~\ref{fig:a100fit} using a polynomial
\begin{equation}
E(A_D) = \sum_{i=0}^{n} a_i \, (A_D - 100)^i
\label{eq:pol}
\end{equation}
and a corresponding formula for the volume integral $J_R$. It has
turned out that the reduced $\chi^2$ of the fit improves when one
increases the number $n$ up to $n = 4$; no further significant
improvement is found for larger values of $n$. These fourth-order
polynomials for $E$ and $J_R$ are shown as lines in
Fig.~\ref{fig:a100fit}. The resulting numbers for $A_D = 100$, i.e.\
the $^{104}$Te $\rightarrow$ $^{100}$Sn \ald , are $E = 5.379$\,MeV
and $J_R = 304.4$\,MeV\,fm$^3$ which is within the error bars of the
values derived above from the neighboring potentials.

Because of the higher \ald\ energy derived in this work, the \ald\
half-life of $^{104}$Te is about a factor of 10 shorter compared to
the predictions of \cite{Xu06}. Experimental data are required to
distinguish between the predictions of this work and Ref.~\cite{Xu06}.

The results for $^{108}$Xe roughly agree with the predictions in
\cite{Xu06}: Xu {\it et al.}\ predict the \ald\ energy $E = 4.44$\,MeV
compared to $E = 4.65 \pm 0.15$\,MeV in this work, and the predicted
half-life in \cite{Xu06} is between 150 and 290\,$\mu$s which should
be compared to the predicted half-life of \thpre\ $= 236$\,$\mu$s
derived from the lower limit $E = 4.5$\,MeV of the energy with $P =
5$\,\%.

\section{Comparison to mass formulae}
\label{sec:mass}
The \ald\ energies of the folding calculation may be compared to
predictions from global mass formulae. Here I restrict myself to the
three selected mass formulae of the so-called Reference Input Parameter
Library RIPL-2 of the IAEA \cite{RIPL2} which are the Finite Range
Droplet Model (FRDM) \cite{Mol95}, the Hartree-Fock-Bogoliubov (HFB)
method \cite{Sam01} in the versions of \cite{RIPL2} and its latest
update \cite{Gor05}, and the simple 10-parameter formula of
Duflo and Zuker (DZ) \cite{Duf95}. The results are listed in Table
\ref{tab:mass}. 
\begin{table}
\setlength{\tabcolsep}{1.5ex}
\caption{ 
Comparison of \ald\ energies from a local extrapolation using folding
potentials (this work) to predictions of global mass formula
\cite{RIPL2,Mol95,Sam01,Gor05,Duf95}. All energies are given in MeV.
}
\label{tab:mass}       
\begin{tabular}{ccrrrr}
\hline\noalign{\smallskip}
& exp.\ or & FRDM & HFB-1 & HFB-2 & DZ \\
& this work & \cite{Mol95} & \cite{Sam01,RIPL2} & \cite{Gor05} &
\cite{Duf95} \\ 
\noalign{\smallskip}\hline\noalign{\smallskip}
$^{104}$Te & $5.42 \pm 0.07^a$ & 6.12 & 4.85 & 4.68 & 5.24 \\
$^{105}$Te & $4.89$            & 6.31 & 4.91 & 4.28 & 4.91 \\
$^{106}$Te & $4.29$            & 6.01 & 4.72 & 4.16 & 4.60 \\
$^{108}$Xe & $4.65 \pm 0.15^a$ & 5.53 & 4.69 & 4.38 & 4.93 \\
$^{109}$Xe & $4.22^b$          & 4.81 & 4.23 & 4.03 & 4.62 \\
$^{110}$Xe & $3.89$            & 4.61 & 3.60 & 3.71 & 4.33 \\
\noalign{\smallskip}\hline
\multicolumn{6}{l}{$^a$ predicted from folding potential} \\
\multicolumn{6}{l}{$^b$ from ground state in $^{109}$Xe to ground
  state in $^{105}$Te} \\
\end{tabular}
\end{table}

The FRDM predictions seem to overestimate the experimental \ald\
energies slightly, especially when approaching the doubly-magic core
$^{100}$Sn. The predictions of HFB-1 and HFB-2 are close to the
experimental values, and also the simple
10-parameter parametrization DZ is in reasonable agreement with the
data. The predictions from the folding potential calculation for
$^{104}$Te and $^{108}$Xe are close to the average values of the above
global mass models \cite{RIPL2,Mol95,Sam01,Gor05,Duf95}.

\section{Accuracy of semi-classical half-lives}
\label{sec:acc}
The results which are presented in Table \ref{tab:halflife} and
Fig.~\ref{fig:pre} have been obtained using the semi-classical
approximation of Eq.~(\ref{eq:gamma}) for the decay width
$\Gamma_\alpha$. From a fully quantum-mechanical analysis the decay
width $\Gamma_\alpha$ is related to the energy dependence of a
resonant scattering phase shift $\delta_L(E)$ by
\begin{equation}
\delta_L(E) = \arctan{\frac{\Gamma_\alpha}{2(E_R-E)}}
\label{eq:phase}
\end{equation}
In practice, it is difficult to determine widths of the order of
1\,$\mu$eV at energies of the order of several MeV because of
numerical problems. For the system $^{104}$Te = $^{100}$Sn $\otimes$
\al\ such an analysis is possible at the limits of numerical
stability. 

\begin{figure}
\resizebox{0.5\textwidth}{!}{
  \includegraphics[55,60][490,265]{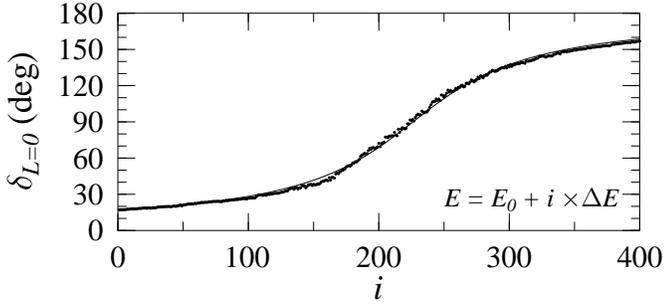}
}
\caption{
  Phase shift $\delta_L$ for the $L = 0$ partial wave for the system
  $^{104}$Te = $^{100}$Sn $\otimes$ \al . The derived width from
  Eq.~(\ref{eq:phase}) is $\Gamma = 1.36 \times 10^{-12}$\,MeV. Note
  the extremely small stepsize of the calculation of $\Delta E = 1.0
  \times 10^{-14}$\,MeV! See text for details.
}
\label{fig:phase}
\end{figure}

In Fig.~\ref{fig:phase} the resonant behavior of the s-wave
phase shift $\delta_{L=0}(E)$ is shown around the resonance energy
$E_R = 5.481$\,MeV which is obtained in the potential with $J_R =
303.76$\,MeV (see Sect.~\ref{sec:predict}). The dots are obtained from
solving the Schr\"odinger equation at $E = E_0 + i \times \Delta E$
with $E_0 = 5.481305851985$\,MeV and $\Delta E = 10^{-14}$\,MeV. The
full line is a fit of data using Eq.~(\ref{eq:phase}) where the
resonance energy $E_R$ and the width $\Gamma_\alpha$ have been
adjusted. This yields $\Gamma_\alpha = 1.36$\,$\mu$eV and a
corresponding half-life of \thcalc\  $= 0.336$\,ns. The semi-classical
approximation in Eq.~(\ref{eq:gamma}) gives \thcalc\ $= 0.313$\,ns
which is about 8\,\% lower than the value from the fully
quantum-mechanical calculation. 

The validity of the semi-classical approximation for $\Gamma_\alpha$ in
Eq.~(\ref{eq:gamma}) is confirmed for the \ald\ of $^{104}$Te by the
above analysis of the scattering phase shift $\delta_L(E)$ with an
uncertainty of less than 10\,\%. For two other nuclei ($^{8}$Be and
$^{212}$Po) the semi-classical approximation deviates by about 30\,\%
from the fully quantum-mechanical value. In all cases the
semi-classical half-life is slightly shorter than the fully
quantum-mechanical result. 

In a detailed study on proton-decay half-lives of proton-rich nuclei
\cite{Abe97} it has been shown that the semi-classical approximation
agrees within about $\pm 10\,\%$ with the result of a direct
calculation of the transition amplitude using the distorted-wave Born
approximation (DWBA) formalism. Surprisingly, the agreement between
the quantum-mechanical DWBA calculation and the semi-classical result
becomes worse in \cite{Abe97} when an improved normalization factor
from  Eq.~(25) of \cite{Abe97} is used compared to the simple
normalization factor in Eq.~(24) of \cite{Abe97} or Eq.~(\ref{eq:f})
in this work. For the case of $^{104}$Te, the \ald\ half-life in the
semi-classical calculation changes from 0.313\,ns using
Eq.~(\ref{eq:f}) to 0.231\,ns using Eq.~(25) of \cite{Abe97}; thus,
the findings in \cite{Abe97} are confirmed.

\section{Properties of $^{104}$Te = $^{100}$Sn $\otimes$ \al }
\label{sec:bound}
From the given potential of the system $^{104}$Te = $^{100}$Sn
$\otimes$ \al\ it is not only possible to determine the \ald\
half-life of the ground state. Following the formalism in
\cite{Buc95}, energies and electromagnetic decay properties of excited
states in $^{104}$Te can be predicted.

The ground state wave function of $^{104}$Te is characterized by $Q =
2N + L = 16$, see Eq.~(\ref{eq:wild}). Further members of this $Q = 16$
band are expected with $J^\pi = 2^+, 4^+, \ldots , 16^+$. It has been
observed that the potential strength parameter $\lambda$ has to be varied
slightly to obtain an excellent prediction of the excitation
energies:
\begin{equation}
\lambda(L) = \lambda(L=0) - c \times L
\label{eq:ohkubo}
\end{equation}
with the constant $c \approx (3-5) \times 10^{-3}$ for neighboring
$N = 50$ $\otimes$ \al\ nuclei 
$^{94}$Mo = $^{90}$Zr $\otimes$ \al\ \cite{Ohk95,Mic98}, 
$^{93}$Nb = $^{89}$Y $\otimes$ \al\ \cite{Kiss06}, 
neighboring $Z = 50$ nuclei
$^{116}$Te = $^{112}$Sn $\otimes$ \al ,
and the systems 
$^{20}$Ne = $^{16}$O $\otimes$ \al\ \cite{Abe93}, 
$^{44}$Ti = $^{40}$Ca $\otimes$ \al\ \cite{Atz96}, and
$^{212}$Po = $^{208}$Pb $\otimes$ \al\ \cite{Hoy94}.

For the following analysis I adopt $\lambda = 1.1005$ which
corresponds to $J_R = 304.29$\,MeV\,fm$^3$ from the average of the two
neighboring systems $^{105,106}$Te = $^{101,102}$Sn $\otimes$ \al\ and
$c = (4.5 \pm 0.3) \times 10^{-3}$ from the neighboring nuclei
$^{93}$Nb, $^{94}$Mo, and $^{116}$Te above $N = 50$ or $Z = 50$
cores. Because the predicted excitation energies $E_x = E - E(0^+)$
(see Table \ref{tab:bound}) are relative to the ground state energy,
the excitation energies do not change significantly when one varies
$\lambda(L=0)$ or $J_R$ within the given uncertainties.
\begin{table}
\setlength{\tabcolsep}{1.5ex}
\caption{ Excitation energies $E_x = E - E(0^+)$ of excited states in
$^{104}$Te = $^{100}$Sn $\otimes$ \al\ woth $Q = 16$, 17, and 18.  
}
\label{tab:bound}       
\begin{tabular}{ccccrrrrrr}
\hline\noalign{\smallskip}
$J^\pi$ & $Q$ & $N$ & $L$ & \multicolumn{1}{c}{$\lambda$} & $E$ (keV)
& $E_x$ (keV) \\ 
\noalign{\smallskip}\hline\noalign{\smallskip}
 $0^+$ & 16 & 8 &  0 & 1.1005 &  5354.2 &    0.0 \\
 $2^+$ & 16 & 7 &  2 & 1.0915 &  6003.5 &  649.3 \\
 $4^+$ & 16 & 6 &  4 & 1.0825 &  6739.6 & 1385.4 \\
 $6^+$ & 16 & 5 &  6 & 1.0735 &  7565.2 & 2211.0 \\
 $8^+$ & 16 & 4 &  8 & 1.0645 &  8477.2 & 3123.0 \\
$10^+$ & 16 & 3 & 10 & 1.0555 &  9469.2 & 4115.0 \\
$12^+$ & 16 & 2 & 12 & 1.0465 & 10543.2 & 5189.0 \\
$14^+$ & 16 & 1 & 14 & 1.0375 & 11731.2 & 6377.0 \\
$16^+$ & 16 & 0 & 16 & 1.0285 & 13097.1 & 7742.9 \\
 $1^-$ & 17 & 8 &  1 & 1.0960 & 10951.4 & 5597.2 \\
 $0^+$ & 18 & 9 &  0 & 1.1005 & $\approx 15$\,MeV $^a$ & $\approx 10$\,MeV $^a$ \\
\noalign{\smallskip}\hline
\multicolumn{6}{l}{$^a$ very broad} \\
\end{tabular}
\end{table}

The first excited $2^+$ state in $^{104}$Te is found at $E_x =
649$\,keV. From the uncertainty of the constant $c$ in
Eq.~(\ref{eq:ohkubo}) one can derive a very small uncertainty for the
potential strength $\lambda(L=2)$ and a resulting uncertainty of about
40\,keV for the excitation energy $E_x$ for the first $2^+$ state. 
Somewhat larger uncertainties are found for $\lambda(L>2)$;
consequently, the uncertainty of the predicted excitation energies
increases up to about 400\,keV for the $16^+$ state at $E_x = 8.55$\,MeV.

In addition, the $1^-$ and $0^+$ band heads of the bands with $Q = 17$
and $Q = 18$ are predicted at energies around $E_x = 5.60$\,MeV and
about 10\,MeV. The $0^+$ state is very broad. It is difficult to
estimate the uncertainty of the predicted energies of the $1^-$ and
$0^+$ states with $Q = 17$ and $Q = 18$ because usually the potential
strength has to be slightly readjusted to obtain a good description of
such bands. A rough estimate for the uncertainty is about 1\,MeV which
corresponds to an uncertainty of about 2\,\% for the potential
strength parameter $\lambda$.

Following the formalism of Ref.\ \cite{Buc95}, reduced transition strengths
of 10.1\,W.u., 14.0\,W.u., and 14.1\,W.u.\ are calculated for the
$2^+ \rightarrow 0^+$, $4^+ \rightarrow 2^+$, and
$6^+ \rightarrow 4^+$ transitions in $^{104}$Te. The corresponding
radiation widths $\Gamma_\gamma$ are slightly larger than the direct
\ald\ widths from the excited states in $^{104}$Te to the ground state
in $^{100}$Sn. The $\gamma$-decay branching ratio
\begin{equation}
b_\gamma = \frac{\Gamma_\gamma}{\Gamma_\gamma +
  \Gamma_\alpha^{\rm{pre}}}
\label{eq:branch}
\end{equation}
is between 86\,\% and 93\,\% for the $2^+$ state, between 62\,\% and
76\,\% for the $4^+$ state, and between 48\,\% and 62\,\% for the
$6^+$ state. This is an extremely important result for future
experiments! If the $\gamma$-decay branch $b_\gamma$ of the first $2^+$
state were small (e.g., of the order of a few per cent), it would be
extremely difficult to produce $^{104}$Te in its ground state
because $^{104}$Te produced in excited states could directly decay
to the $^{100}$Sn ground state by \al\ emission.

It is interesting to note that the predicted branchings $b_\gamma$ are
not very sensitive to the predicted excitation energy. E.g., if the
excitation energy of the first excited $2^+$ state in $^{104}$Te is
$E_x = 1$\,MeV, the radiation width $\Gamma_\gamma$ increases with
$E_\gamma^5$ by a factor of about 9 and the width $\Gamma_\alpha$
increases by a factor of about 8 because of the reduced Coulomb
barrier. Thus, $b_\gamma$ values close to unity are very
likely. Consequently, a direct production reaction like e.g.\
$^{50}$Cr($^{58}$Ni,4$n$)$^{104}$Te similar to the experiment in
\cite{Sew06} should be feasible. However, only the indirect production
via the \ald\ of $^{108}$Xe in a reaction like e.g.\
$^{54}$Fe($^{58}$Ni,4$n$)$^{108}$Xe similar to \cite{Lid06} ensures
the production of $^{104}$Te in its ground state.

\section{Conclusions}
\label{sec:conc}
The systematic properties of folding potentials provide a powerful
tool for the analysis of the system $^{104}$Te = $^{100}$Sn $\otimes$
\al\ above the doubly-magic $^{100}$Sn core. In particular, \ald\
energies and half-lives can be predicted with relatively small
uncertainties. The predicted \ald\ energy for $^{104}$Te is $E = 5.42
\pm 0.07$\,MeV, and the corresponding half-life is \thpre\ $= 5$\,ns
with an uncertainty of a factor of three.

Excitation energies and decay properties of the members of the $Q =
16$ rotational band in $^{104}$Te are calculated, and the predicted
values have small uncertainties. For the first excited $2^+$ state in
$^{104}$Te one obtains $E_x = 650 \pm 40$\,keV. The $\gamma$-decay
strength to the ground state in $^{104}$Te is about 10 Weisskopf
units. The corresponding radiation width $\Gamma_\gamma$ is about a
factor of 10 larger than the \ald\ width $\Gamma_\alpha$ to the ground
state in $^{100}$Sn.

The finding that $\Gamma_\gamma$ is larger than $\Gamma_\alpha$ for
excited states in $^{104}$Te is important for the experimental
production of $^{104}$Te in its ground state and the measurement of
the \ald\ half-life of $^{104}$Te. The condition $\Gamma_\gamma >
\Gamma_\alpha$ allows to use reactions which produce $^{104}$Te in
excited states because these states preferentially decay to the
$^{104}$Te ground state. However, only the indirect production of
$^{104}$Te via the \ald\ of $^{108}$Xe safely guarantees that
$^{104}$Te is produced in its ground state.

\medskip
\noindent
I thank Z.\ Ren, Gy. Gyurky, and Zs.\ F\"ul\"op for encouraging
discussions and the referees for their constructive reports.


\end{document}